\documentclass[reprint,aps,ams,amsmath,prl,twocolumn,longbibliography,superscriptaddress]{revtex4-1}
\usepackage{graphics}
\usepackage{graphicx}
\usepackage{epsfig}
\usepackage{amsmath}
\usepackage{amssymb}
\usepackage{amsfonts}
\usepackage{bm}
\usepackage{color}
\usepackage{bbm}
\usepackage{hyperref}
\usepackage[normalem]{ulem}
\usepackage{comment}
\usepackage{soul}

\begin{document}

\title{Spin drag mechanism of giant thermal magnetoresistance}

\author{Alex Levchenko}
\affiliation{Department of Physics, University of Wisconsin-Madison, Madison, Wisconsin 53706, USA}

\author{A. V. Andreev}
\affiliation{Department of Physics, University of Washington, Seattle, Washington 98195, USA}

\date{July 12, 2024}

\begin{abstract}
We study hydrodynamic thermal transport in high-mobility two-dimensional electron systems placed in an in-plane magnetic field, and identify a new mechanism of thermal magnetotransport. This mechanism is caused by drag between the electron populations with opposite spin polarization, which arises in the presence of a hydrodynamic flow of heat. In high mobility systems, spin drag results in strong thermal magnetoresistance, which becomes of the order of 100 \% at relatively small spin polarization of the electron liquid. We express  the thermal magnetoresistance in terms of intrinsic dissipative coefficients of electron fluid and show that it is primarily determined by the spin diffusion constant.
\end{abstract}

\maketitle

The role of mutual drag in transport phenomena is pivotal to understanding kinetic properties of metals, semiconductors, and insulators. Perhaps the most prominent examples of these effects are phonon drag \cite{PD-review}, Coulomb drag in bilayers \cite{CD-Review}, and magnon drag in magnetic systems \cite{Costache:2012}.

Phonon drag, also known as the Gurevich effect, leads to significant deviations in the thermopower of various materials from the predictions based solely on electronic theory. These deviations are caused by the transfer of momentum from the electrons to the phonons, resulting in a substantial heat flow carried by phonons. Likewise, the momentum transfer induced by interlayer electron collisions, mediated by Coulomb interactions, gives rise to drag resistance \cite{MacDonald:1993,JauhoSmith:1993,Flensberg:1995,KamenevOreg:1995}. This occurs when one layer (e.g. a quantum wire or a two-dimensional electron system) is driven out of equilibrium by a current, inducing a nonlocal voltage in the adjacent layer.

In ferromagnetic metals, electron-magnon scattering  produces thermoelectric anomalies similar to the phonon drag effect \cite{Bailyn:1962,Blatt:1967,Berger:1976,Miura:2012}. In magnetic insulators, one can realize a nonlocal magnon drag induced by magnetic dipolar interactions between the layers \cite{Flatte:2016}. In complete analogy to the Coulomb drag,
a magnon current in one layer induces a chemical potential gradient and/or a temperature gradient in the other layer, which are characterized by the magnon current transresistivity and the magnon thermal transresistivity. The effect of mutual drag between phonons and spin excitations has been also discussed in the literature in the context of the thermal conductivity of a quantum spin system \cite{Chernyshev:2010}.

In this work, we introduce a different drag-induced thermal effect. We show that near charge neutrality, spin polarization of the electrons strongly affects the thermal conductivity.
The salient feature of heat transport at charge neutrality is that it can proceed via the hydrodynamic flow of the neutral electron-hole plasma. We thus focus our consideration on the hydrodynamic regime, which attracted significant attention in recent years, see reviews \cite{Lucas:2018,Levchenko:2020,Narozhny:2022} and references therein. In a pristine system, the thermal conductivity becomes infinite, and thus very sensitive to disorder and other perturbations. In realistic systems, the thermal conductivity is limited by the disorder-induced friction, which can be made sufficiently weak in high mobility systems. Significant hydrodynamic enhancement of thermal conductivity in graphene systems has been reported in Ref.~\cite{Crossno:2016}. Continuing progress in nanofabrication enables fabrication of samples with an even greater mobility, making the thermal conductivity of the system extremely sensitive to other perturbations. Here, we show that thermal transport properties of high mobility systems become very sensitive to spin polarization of the liquid, leading to strong thermal magnetoresistivity. The microscopic mechanism of this phenomenon in the hydrodynamic regime involves spin diffusion (or spin drag). The physical reason for this effect can be understood by observing that since thermal conductivity is measured at zero spin-current, the convective part of the spin current must be compensated by spin-diffusion relative to the liquid. This dramatically increases dissipation at nonzero spin polarization, resulting in giant thermal magnetoresistivity.
Our predictions may enable probing spin drag in electron-hole plasma in graphene by thermal measurements \cite{Waissman:2022,Talanov:2024}.

The hydrodynamic description of electron transport in a solid applies provided the rate of momentum- and energy-conserving electron-electron collisions exceeds the momentum and energy relaxation rates on impurities and phonons \cite{Gurzhi:1968,AKS}. Therefore macroscopic hydrodynamic equations express conservation of the number of particles, energy, and momentum of the electron liquid. In addition, in multivalley conductors or in the spin-polarized systems additional approximate conservation laws are possible for (pseudo-) spin degrees of freedom.

In what follows we consider a two-dimensional system in which the electron fluid is partially spin polarized by an in-plane external magnetic field. We assume that the spin-orbit interaction is absent, so that the spin component along the magnetic field is conserved. Experiments show that even at room temperature spin transport in single-, bi-, and trilayer graphene devices exhibit nanosecond spin lifetimes with spin diffusion lengths reaching 10 $\mu$m \cite{Drogeler:2014,Drogeler:2016}. These observations justify our assumptions.

When the electron system is tuned to the charge neutrality point by an applied gate voltage the hydrodynamic flow is decoupled from charge current.
Thus, the hydrodynamic equations involve only the entropy current density $\bm{j}_s$ and spin current density $\bm{j}_\sigma$. In a steady state and in the linear regime these two currents are conserved, which is expressed by the continuity equation of the form
\begin{equation}\label{eq:J}
\bm{\nabla}\cdot\vec{\bm{J}}=0, \quad \vec{\bm{J}}=\vec{x}\bm{u}-\hat{\Xi}\vec{\bm{X}}.
\end{equation}
Here we used column vector notations
\begin{equation}
\label{eq:column_vector}
\vec{\bm{J}}=\left(\begin{array}{c}
\bm{j}_s \\ \bm{j}_\sigma \end{array}\right),\quad
\vec{x}=\left(\begin{array}{c}
s \\ \varsigma \end{array}\right),\quad \vec{\bm{X}}=\left(\begin{array}{c}\bm{\nabla}T \\ \bm{\nabla}\mu_\sigma\end{array}\right),
\end{equation}
where $s$ and $\varsigma$ are, respectively, entropy and spin densities, while $\vec{\bm{X}}$ is the vector of conjugated thermodynamic forces defined by gradients of temperature $T$ and spin chemical potential $\mu_\sigma$. It should be noted that $\varsigma$ refers to the projection on the axis of the external field, and the other spin components do not appear in the hydrodynamic description because they are not conserved due to spin precession. The first term in Eq. \eqref{eq:J} corresponds to the convective part of the current. It is worthwhile to stress that in the collision-dominated regime hydrodynamic velocity $\bm{u}(\bm{r})$ is the same for all the spin components. This is different from the regime of  spin Coulomb drag  \cite{Vignale:2000,Vignale:2003}, where the populations of spin-up and spin-down electrons have different drift velocities.

The net currents of entropy and spin consist of the  convective currents  produced by the thermally-driven flow of the partially spin-polarized electron liquid, and the dissipative currents relative to the liquid described by the second term in Eq. \eqref{eq:J}. The latter are characterized by the matrix of intrinsic kinetic coefficients
\begin{equation}
\hat{\Xi}=\left(\begin{array}{cc}\kappa/T & \gamma_\sigma/T \\ \gamma_\sigma/T & D_\sigma\end{array}\right),
\end{equation}
which satisfies the  Onsager symmetry principle \cite{Onsager-I,Onsager-II}.
The diagonal elements contributing to dissipation are the intrinsic thermal conductivity $\kappa$ and the spin diffusion constant $D_\sigma$ of the electron liquid. The off-diagonal elements describe the so-called spin Seebeck effect, which has been studied in much detail for ferromagnets in the field of spin caloritronics \cite{Bauer:2012}. Since the spin density is odd under time reversal symmetry and energy is not, the Onsager symmetry requires the intrinsic spin thermoconductivity to be odd function of the magnetic field $\gamma_\sigma(H)=-\gamma_\sigma(-H)$.

In the stationary regime the force balance condition for an element of the fluid can be expressed in the form
\begin{equation}\label{eq:force-balance}
\bm{\nabla}\cdot\Sigma-k\bm{u}=\vec{x}^{\mathbb{T}}\vec{\bm{X}},
\end{equation}
where the first term in the left hand side represents the divergence of the viscous stress tensor \cite{LL-V6}
\begin{equation}
\Sigma_{ij}=\eta(\partial_iu_j+\partial_j u_i)+(\zeta-\eta)\delta_{ij}\partial_k u_k
\end{equation}
with $\eta$ and $\zeta$ being, respectively, shear (first) and bulk (second) viscosity of the electron liquid.
The force density in the right hand side of Eq. \eqref{eq:force-balance} comes from the local gradients of pressure in the fluid $P$. To express it in this form we used the thermodynamic relation $\bm{\nabla}P=s\bm{\nabla}T+\varsigma\bm{\nabla}\mu_\sigma=\vec{x}^{\mathbb{T}}\vec{\bm{X}}$ and the column vector notations of Eq. \eqref{eq:column_vector}.
The superscript $\mathbb{T}$ denotes transposition. The remaining term in Eq.~\eqref{eq:force-balance} describes the generic disorder-induced friction characterized by the friction coefficient $k$. For weak disorder whose correlation   radius $\xi$ exceeds the electron inelastic mean free path $l_{\text{ee}}$, the coefficient of friction $k$ can be expressed in terms of the local density variations $\delta n(\bm{r})$ induced by disorder potential and the intrinsic conductivity $\sigma$ as follows \cite{Li:2020}
\begin{equation}
\label{eq:k_delta_n}
k=\frac{e^2}{2\sigma}\langle\delta n^2\rangle,
\end{equation}
where $\langle\ldots\rangle$ denotes spatial averaging. We recall that the intrinsic conductivity does not vanish in generic electron liquids which do not possess Galilean invariance.
The assumed model is motivated by the experimental observations of the long-range disorder in graphene devices in the form of charge puddles \cite{Yacoby:2008,Crommie:2009,LeRoy:2011} (with the typical scale of $\xi\sim100$ nm). The local form of Eq. \eqref{eq:force-balance} is supported by the recent analysis presented in Refs. \cite{Lucas:2016,Li:2020,Levchenko:2024}, where it was shown that for a weakly-disordered system one can develop an effective renormalized hydrodynamic description on length scales exceeding $\xi$ \footnote{It has been shown in Ref.~\cite{Aleiner:2009} that for $e^2/\hbar v\ll 1$ the relaxation length $l_{\text{rec}}$ associated with electron-hole recombination is parametrically longer than the electron-electron scattering length $l_{\text{ee}}$. In this case, the hydrodynamic description applies if the disorder correlation radius exceeds $l_{\text{rec}}$. However, in graphene devices, where the interaction constant is of order unity, $l_{\text{rec}}$ and $l_{\text{ee}}$ are not expected to be parametrically different, and the condition $\xi \gg l_{\text{ee}}$ is adequate for the applicability of the hydrodynamic description.}.

For a given geometry of the sample and appropriate boundary conditions Eqs. \eqref{eq:J} and \eqref{eq:force-balance} uniquely determine the flow profile.
The precise form of macroscopic transport coefficients follows from the expression for the entropy production rate \footnote{See file of supplemental materials for further details}
\begin{equation}\label{eq:dotS}
T\dot{S}=\left\langle\Sigma_{ij}\partial_ju_i+\vec{\bm{X}}^{\mathbb{T}}\hat{\Xi}\vec{\bm{X}}+k\bm{u}^2\right\rangle
\end{equation}
that should be equated to the Joule heating power $\mathcal{P}=\vec{\bm{J}}^{\mathbb{T}}\hat{\mathcal{R}}\vec{\bm{J}}$. The matrix elements of $\hat{\mathcal{R}}$ define thermal and spin resistances. Alternatively, one can proceed via the linear response to relate currents to applied gradients and thus infer the effective matrix of conductivities whose inverse is $\hat{\mathcal{R}}$.
Below we use the second route as it is more straightforward for the problem at hand.

The macroscopic thermal conductivity $\varkappa$ is defined as the proportionality coefficient between the entropy current and the temperature gradient at vanishing spin current
\begin{equation}\label{eq:kappa}
\varkappa=-T(\bm{j}_s/\bm{\nabla} T)_{\bm{j}_\sigma=0}.
\end{equation}
In the absence of spin polarization, the thermal conductivity of large systems at charge neutrality is determined by the friction coefficient and is independent of the liquid viscosity~\cite{Crossno:2016,Lucas:2016,Li:2020}. We show below that for spin-polarized systems the thermal resistivity remains independent from viscosity and is controlled by the spin diffusion coefficient.
To this end, we notice that the comparison of the gradient terms in Eq. \eqref{eq:force-balance} describing viscous stresses, $\bm{\nabla}\cdot\Sigma=(\eta+\zeta)\bm{\nabla}^2\bm{u}$, to the friction term, $k\bm{u}$, introduces a characteristic length scale in the problem, which is the Gurzhi length \cite{Gurzhi:1968}
\begin{equation}
l_{\text{G}}=\sqrt{\frac{\eta+\zeta}{k}}.
\end{equation}
Therefore, if the sample size $L$ is smaller than $l_{\text{G}}$ the flow profile is essentially inhomogeneous (Poiseuille-like) and thus viscous effects play an important role. In the opposite case of wide devices, $L\gg l_{\text{G}}$, the flow is mostly uniform except in the boundary layer of thickness $\sim l_{\text{G}}$ near the sample edges. Based on this reasoning we assume the following hierarchy of length scales $\xi \ll l_{\text{G}}\ll L$. In this limit we may neglect the gradient terms in Eq. \eqref{eq:force-balance} in the  bulk of the sample,  which significantly simplifies the consideration. Then trivially solving for $\bm{u}$ we find $\bm{u}=-(\vec{x}^{\mathbb{T}}\vec{\bm{X}})/k$. At the same time, the required condition on the vanishing spin current gives us from Eq. \eqref{eq:J} that $\bm{u}=\frac{D_\sigma}{\varsigma}\bm{\nabla}\mu_\sigma+\frac{\gamma_\sigma}{\varsigma T}\bm{\nabla}T$. These two equations fix $\bm{u}$ and give a local relationship between $\bm{\nabla}\mu_\sigma$ and $\bm{\nabla}T$
\begin{equation}
\bm{\nabla}\mu_\sigma=-\bm{\nabla}T\frac{s+\frac{\gamma_\sigma k}{\varsigma T}}{\varsigma+\frac{D_\sigma k}{\varsigma}}.
\end{equation}
Having determined both $\bm{u}$ and $\bm{\nabla}\mu_\sigma$ in terms of $\bm{\nabla}T$, we insert both expressions into the first row of Eq. \eqref{eq:J}, which gives us the entropy current in the presence of the thermal spin drag. After straightforward algebra we obtain the following result for the effective thermal conductivity from Eq. \eqref{eq:kappa}
\begin{equation}\label{eq:thMR}
\varkappa(H)=\kappa+T\frac{\left(\frac{sD_\sigma}{\varsigma}-\frac{\gamma_\sigma}{T}\right)\left(s+\frac{\gamma_\sigma k}{T\varsigma}\right)}{\varsigma+\frac{D_\sigma k}{\varsigma}}-\frac{s\gamma_\sigma}{\varsigma}.
\end{equation}
For small spin polarizations it is safe to assume that $s\gg \mathrm{max}\{\frac{\varsigma\gamma_\sigma}{TD_\sigma},\frac{k\gamma_\sigma}{\varsigma T}\}$, so that only $s^2D_\sigma/\varsigma$ should be retained in the numerator of the second term of Eq. \eqref{eq:thMR}, and the last term can be dropped as well. Indeed, for example, for the graphene monolayer $s\sim(T/v)^2$, where $v$ is the band velocity of graphene. For long range disorder we have $\xi\gg l_T\equiv v/T$. Therefore, the above conditions are satisfied in the hydrodynamic regime. Furthermore, since for weak disorder $\kappa\ll Ts^2/k$,  our main result can be simplified to
\begin{equation}\label{eq:varkappa}
\varkappa(H)\approx T\frac{s^2D_\sigma}{kD_\sigma+\varsigma^2}.
\end{equation}

Note that in the absence of spin diffusion the thermal conductivity in Eq.~\eqref{eq:varkappa} vanishes. This occurs because a hydrodynamic flow of spin-polarized liquid at vanishing spin current is impossible in the absence of spin diffusion. Thus, in the ideal fluid limit, where both intrinsic thermal conductivity and spin diffusion coefficient vanish, the system becomes a thermal insulator. This corresponds to spin-induced stagnation of the electron liquid, which may be used to create spin-actuated thermal valves. A similar stagnation effect arises in hydrodynamic transport of charge away from charge neutrality. In that case simultaneous conservation of currents of charge, entropy, and  (for a partially spin-polarized liquid) spin precludes potential flow of an ideal liquid in a smooth external potential~\cite{AKS,Skinner:2024}, resulting in diverging resistivity of 1D systems in the ideal fluid limit  \cite{Levchenko:2010,DeGottardi:2015}.

We note that the reduction of the thermal magnetoconductivity by spin polarization reaches $\sim $ 100\%  when the spin density $\varsigma(H)$ becomes of the order of root mean square of the charge density fluctuations induced by disorder, namely when $\varsigma\sim\sqrt{\frac{D_\sigma e^2}{\sigma}}\sqrt{\langle\delta n^2\rangle}$. At such weak fields magnetic field dependence of the spin diffusion constant $D_\sigma(H)$ and intrinsic conductivity $\sigma(H)$ can be neglected. Furthermore, equation \eqref{eq:varkappa} remains valid even in the case when spin polarization arises due to spontaneous symmetry breaking as long as the hydrodynamic limit can still be justified.

In the case of field-induced spin polarization Eq.~\eqref{eq:varkappa} can be used to obtain thermal magnetoconductivity at low magnetic fields. Indeed,
we write the spin density in the form $\varsigma=\chi H$, where $\chi$ denotes the spin susceptibility. In this case Eq. \eqref{eq:varkappa} yields a  Lorentzian dependence of thermal conductivity on $H$,
\begin{equation}
\varkappa(H)\approx\frac{Ts^2}{k}\frac{1}{1+(H/H_\sigma)^2},\quad H_\sigma=\frac{\sqrt{kD_\sigma}}{\chi}.
\end{equation}
The corresponding thermal resistivity $\varrho_{\text{th}}=\varkappa^{-1}$ is thus positive and quadratic. It is of interest to note that the relative thermal magnetoresistance,
\begin{equation}\label{eq:rel-th-MR}
\Delta\varrho_{\text{th}}(H)\equiv\frac{\varrho_\text{th}(H)-\varrho_\text{th}(0)}{\varrho_\text{th}(0)}=\frac{\varsigma^2}{kD_\sigma},
\end{equation}
provides a way to extract the spin diffusion coefficient $D_\sigma$ from thermal transport measurements, since the degree of spin polarization and the strength of disorder can be determined from the independent experimental probes. At stronger field, when spin density saturates, the resistivity $\varrho_{\text{th}}$ also saturates to a constant value. The effect is anomalously strong since $\Delta\varrho_{\text{th}}\sim 1$ already for $H\sim H_\sigma$.
 
To evaluate magneto-thermal conductivity induced by spin drag away from charge neutrality it is important to realize that the right-hand-side of Eq. \eqref{eq:kappa} must be evaluated at vanishing spin and electric current. 
Although the latter condition is no longer automatically satisfied at nonvanishing particle density $n$, for  $n \ll s$ the result can be obtained by a straightforward extension of the above consideration. This requires two main modifications. First, the Coulomb force $-en\bm{E}$, where the electromotive force (EMF) $e \bm{E}$ is the gradient of the local electrochemical potential, must be added to the right-hand-side of the force balance condition, Eq. \eqref{eq:force-balance}. Second, the constitutive relations in Eq. \eqref{eq:J} have to be augmented to include the following expression for the electric current density, $\bm{j}_e=en\bm{u}+\sigma\bm{E}-\frac{\gamma}{T}\bm{\nabla}T$, where $\gamma$ is the intrinsic thermoelectric coefficient.  At $n\ll s$  the thermoelectric contribution to $\bm{j}_{e}$ may be neglected because $\gamma/T\sim n/s\ll1$. In this case,  the condition of vanishing current
 gives  $\bm{E}=-(en/\sigma)\bm{u}$. Inserting this expression into the force-balance condition gives a force term $(en)^2\bm{u}/\sigma$, which amounts  to the following redefinitoin of the effective friction coefficient in the previous expressions, 
\begin{equation}
k\to k(n)=k+\frac{e^2}{\sigma}n^2.
\end{equation}
Thus, the dependence of
the relative thermal magnetoresistance in Eq. \eqref{eq:rel-th-MR} on $n$ has the form of the Lorentzian with a width  $\sim \sqrt{\langle\delta n^2\rangle}$, as follows from Eq.~\eqref{eq:k_delta_n}.
 The limit of charge neutrality corresponds to $n\ll\delta n$. In modern monolayer graphene devices density inhomogeneity ranges between $\sqrt{\langle\delta n^2\rangle}\sim (5\div10)\times 10^{9}$ cm$^{-2}$ \cite{Crossno:2016,Talanov:2024,Berdyugin:2023}. This is still safely compatible with the condition $\{n,\sqrt{\langle\delta n^2\rangle}  \}<s$ at $T\sim 100$ K in monolayer graphene where hydrodynamic behavior of Dirac plasma is expected to emerge. 

In closing, we note that our consideration focused on the bulk contribution to thermal spin drag magnetotransport where momentum relaxation is driven by the disorder potential and the hydrodynamic flow velocity is uniform. In devices whose dimensions are smaller or comparable to the Gurzhi length there will be additional contribution to the thermal resistance, which is  determined by the viscous flow near sample boundaries. An extension of the present theory to the devices with Hall bar and Corbino geometry will be presented elsewhere.

We thank Philip Kim for useful discussion. This research project was financially supported by the National Science Foundation Grant No. DMR-2203411 and H. I. Romnes Faculty Fellowship provided by the University of Wisconsin-Madison Office of the Vice Chancellor for Research and Graduate Education with funding from the Wisconsin Alumni Research Foundation (A. L.). A. V. A. acknowledges support from the National Science Foundation through the MRSEC Grant No. DMR-1719797, the Thouless Institute for Quantum Matter (TIQM), and the
College of Arts and Sciences at the University of Washington.

\bibliography{biblio}

\begin{thebibliography}{42}%
\makeatletter
\providecommand \@ifxundefined [1]{%
 \@ifx{#1\undefined}
}%
\providecommand \@ifnum [1]{%
 \ifnum #1\expandafter \@firstoftwo
 \else \expandafter \@secondoftwo
 \fi
}%
\providecommand \@ifx [1]{%
 \ifx #1\expandafter \@firstoftwo
 \else \expandafter \@secondoftwo
 \fi
}%
\providecommand \natexlab [1]{#1}%
\providecommand \enquote  [1]{``#1''}%
\providecommand \bibnamefont  [1]{#1}%
\providecommand \bibfnamefont [1]{#1}%
\providecommand \citenamefont [1]{#1}%
\providecommand \href@noop [0]{\@secondoftwo}%
\providecommand \href [0]{\begingroup \@sanitize@url \@href}%
\providecommand \@href[1]{\@@startlink{#1}\@@href}%
\providecommand \@@href[1]{\endgroup#1\@@endlink}%
\providecommand \@sanitize@url [0]{\catcode `\\12\catcode `\$12\catcode
  `\&12\catcode `\#12\catcode `\^12\catcode `\_12\catcode `\%12\relax}%
\providecommand \@@startlink[1]{}%
\providecommand \@@endlink[0]{}%
\providecommand \url  [0]{\begingroup\@sanitize@url \@url }%
\providecommand \@url [1]{\endgroup\@href {#1}{\urlprefix }}%
\providecommand \urlprefix  [0]{URL }%
\providecommand \Eprint [0]{\href }%
\providecommand \doibase [0]{http://dx.doi.org/}%
\providecommand \selectlanguage [0]{\@gobble}%
\providecommand \bibinfo  [0]{\@secondoftwo}%
\providecommand \bibfield  [0]{\@secondoftwo}%
\providecommand \translation [1]{[#1]}%
\providecommand \BibitemOpen [0]{}%
\providecommand \bibitemStop [0]{}%
\providecommand \bibitemNoStop [0]{.\EOS\space}%
\providecommand \EOS [0]{\spacefactor3000\relax}%
\providecommand \BibitemShut  [1]{\csname bibitem#1\endcsname}%
\let\auto@bib@innerbib\@empty
\bibitem [{\citenamefont {Fletcher}\ \emph {et~al.}(2003)\citenamefont
  {Fletcher}, \citenamefont {Zaremba},\ and\ \citenamefont
  {Zeitler}}]{PD-review}%
  \BibitemOpen
  \bibfield  {author} {\bibinfo {author} {\bibfnamefont {R.}~\bibnamefont
  {Fletcher}}, \bibinfo {author} {\bibfnamefont {E.}~\bibnamefont {Zaremba}}, \
  and\ \bibinfo {author} {\bibfnamefont {U.}~\bibnamefont {Zeitler}},\
  }\bibfield  {title} {\enquote {\bibinfo {title} {{Phonon drag thermopower of
  low-dimensional systems}},}\ }in\ \href {\doibase
  10.1093/acprof:oso/9780198507321.003.0005} {\emph {\bibinfo {booktitle}
  {{Electron-Phonon Interactions in Low-Dimensional Structures}}}}\ (\bibinfo
  {publisher} {Oxford University Press},\ \bibinfo {year} {2003})\BibitemShut
  {NoStop}%
\bibitem [{\citenamefont {Narozhny}\ and\ \citenamefont
  {Levchenko}(2016)}]{CD-Review}%
  \BibitemOpen
  \bibfield  {author} {\bibinfo {author} {\bibfnamefont {B.~N.}\ \bibnamefont
  {Narozhny}}\ and\ \bibinfo {author} {\bibfnamefont {A.}~\bibnamefont
  {Levchenko}},\ }\bibfield  {title} {\enquote {\bibinfo {title} {Coulomb
  drag},}\ }\href {\doibase 10.1103/RevModPhys.88.025003} {\bibfield  {journal}
  {\bibinfo  {journal} {Rev. Mod. Phys.}\ }\textbf {\bibinfo {volume} {88}},\
  \bibinfo {pages} {025003} (\bibinfo {year} {2016})}\BibitemShut {NoStop}%
\bibitem [{\citenamefont {Costache}\ \emph {et~al.}(2012)\citenamefont
  {Costache}, \citenamefont {Bridoux}, \citenamefont {Neumann},\ and\
  \citenamefont {Valenzuela}}]{Costache:2012}%
  \BibitemOpen
  \bibfield  {author} {\bibinfo {author} {\bibfnamefont {Marius~V.}\
  \bibnamefont {Costache}}, \bibinfo {author} {\bibfnamefont {German}\
  \bibnamefont {Bridoux}}, \bibinfo {author} {\bibfnamefont {Ingmar}\
  \bibnamefont {Neumann}}, \ and\ \bibinfo {author} {\bibfnamefont {Sergio~O.}\
  \bibnamefont {Valenzuela}},\ }\bibfield  {title} {\enquote {\bibinfo {title}
  {Magnon-drag thermopile},}\ }\href {\doibase 10.1038/nmat3201} {\bibfield
  {journal} {\bibinfo  {journal} {Nature Materials}\ }\textbf {\bibinfo
  {volume} {11}},\ \bibinfo {pages} {199--202} (\bibinfo {year}
  {2012})}\BibitemShut {NoStop}%
\bibitem [{\citenamefont {Zheng}\ and\ \citenamefont
  {MacDonald}(1993)}]{MacDonald:1993}%
  \BibitemOpen
  \bibfield  {author} {\bibinfo {author} {\bibfnamefont {Lian}\ \bibnamefont
  {Zheng}}\ and\ \bibinfo {author} {\bibfnamefont {A.~H.}\ \bibnamefont
  {MacDonald}},\ }\bibfield  {title} {\enquote {\bibinfo {title} {Coulomb drag
  between disordered two-dimensional electron-gas layers},}\ }\href {\doibase
  10.1103/PhysRevB.48.8203} {\bibfield  {journal} {\bibinfo  {journal} {Phys.
  Rev. B}\ }\textbf {\bibinfo {volume} {48}},\ \bibinfo {pages} {8203--8209}
  (\bibinfo {year} {1993})}\BibitemShut {NoStop}%
\bibitem [{\citenamefont {Jauho}\ and\ \citenamefont
  {Smith}(1993)}]{JauhoSmith:1993}%
  \BibitemOpen
  \bibfield  {author} {\bibinfo {author} {\bibfnamefont {Antti-Pekka}\
  \bibnamefont {Jauho}}\ and\ \bibinfo {author} {\bibfnamefont {Henrik}\
  \bibnamefont {Smith}},\ }\bibfield  {title} {\enquote {\bibinfo {title}
  {Coulomb drag between parallel two-dimensional electron systems},}\ }\href
  {\doibase 10.1103/PhysRevB.47.4420} {\bibfield  {journal} {\bibinfo
  {journal} {Phys. Rev. B}\ }\textbf {\bibinfo {volume} {47}},\ \bibinfo
  {pages} {4420--4428} (\bibinfo {year} {1993})}\BibitemShut {NoStop}%
\bibitem [{\citenamefont {Flensberg}\ \emph {et~al.}(1995)\citenamefont
  {Flensberg}, \citenamefont {Hu}, \citenamefont {Jauho},\ and\ \citenamefont
  {Kinaret}}]{Flensberg:1995}%
  \BibitemOpen
  \bibfield  {author} {\bibinfo {author} {\bibfnamefont {Karsten}\ \bibnamefont
  {Flensberg}}, \bibinfo {author} {\bibfnamefont {Ben Yu-Kuang}\ \bibnamefont
  {Hu}}, \bibinfo {author} {\bibfnamefont {Antti-Pekka}\ \bibnamefont {Jauho}},
  \ and\ \bibinfo {author} {\bibfnamefont {Jari~M.}\ \bibnamefont {Kinaret}},\
  }\bibfield  {title} {\enquote {\bibinfo {title} {Linear-response theory of
  {C}oulomb drag in coupled electron systems},}\ }\href {\doibase
  10.1103/PhysRevB.52.14761} {\bibfield  {journal} {\bibinfo  {journal} {Phys.
  Rev. B}\ }\textbf {\bibinfo {volume} {52}},\ \bibinfo {pages} {14761--14774}
  (\bibinfo {year} {1995})}\BibitemShut {NoStop}%
\bibitem [{\citenamefont {Kamenev}\ and\ \citenamefont
  {Oreg}(1995)}]{KamenevOreg:1995}%
  \BibitemOpen
  \bibfield  {author} {\bibinfo {author} {\bibfnamefont {Alex}\ \bibnamefont
  {Kamenev}}\ and\ \bibinfo {author} {\bibfnamefont {Yuval}\ \bibnamefont
  {Oreg}},\ }\bibfield  {title} {\enquote {\bibinfo {title} {Coulomb drag in
  normal metals and superconductors: Diagrammatic approach},}\ }\href {\doibase
  10.1103/PhysRevB.52.7516} {\bibfield  {journal} {\bibinfo  {journal} {Phys.
  Rev. B}\ }\textbf {\bibinfo {volume} {52}},\ \bibinfo {pages} {7516--7527}
  (\bibinfo {year} {1995})}\BibitemShut {NoStop}%
\bibitem [{\citenamefont {Bailyn}(1962)}]{Bailyn:1962}%
  \BibitemOpen
  \bibfield  {author} {\bibinfo {author} {\bibfnamefont {M.}~\bibnamefont
  {Bailyn}},\ }\bibfield  {title} {\enquote {\bibinfo {title} {Maximum
  variational principle for conduction problems in a magnetic field, and the
  theory of magnon drag},}\ }\href {\doibase 10.1103/PhysRev.126.2040}
  {\bibfield  {journal} {\bibinfo  {journal} {Phys. Rev.}\ }\textbf {\bibinfo
  {volume} {126}},\ \bibinfo {pages} {2040--2054} (\bibinfo {year}
  {1962})}\BibitemShut {NoStop}%
\bibitem [{\citenamefont {Blatt}\ \emph {et~al.}(1967)\citenamefont {Blatt},
  \citenamefont {Flood}, \citenamefont {Rowe}, \citenamefont {Schroeder},\ and\
  \citenamefont {Cox}}]{Blatt:1967}%
  \BibitemOpen
  \bibfield  {author} {\bibinfo {author} {\bibfnamefont {F.~J.}\ \bibnamefont
  {Blatt}}, \bibinfo {author} {\bibfnamefont {D.~J.}\ \bibnamefont {Flood}},
  \bibinfo {author} {\bibfnamefont {V.}~\bibnamefont {Rowe}}, \bibinfo {author}
  {\bibfnamefont {P.~A.}\ \bibnamefont {Schroeder}}, \ and\ \bibinfo {author}
  {\bibfnamefont {J.~E.}\ \bibnamefont {Cox}},\ }\bibfield  {title} {\enquote
  {\bibinfo {title} {Magnon-drag thermopower in iron},}\ }\href {\doibase
  10.1103/PhysRevLett.18.395} {\bibfield  {journal} {\bibinfo  {journal} {Phys.
  Rev. Lett.}\ }\textbf {\bibinfo {volume} {18}},\ \bibinfo {pages} {395--396}
  (\bibinfo {year} {1967})}\BibitemShut {NoStop}%
\bibitem [{\citenamefont {Grannemann}\ and\ \citenamefont
  {Berger}(1976)}]{Berger:1976}%
  \BibitemOpen
  \bibfield  {author} {\bibinfo {author} {\bibfnamefont {G.~N.}\ \bibnamefont
  {Grannemann}}\ and\ \bibinfo {author} {\bibfnamefont {L.}~\bibnamefont
  {Berger}},\ }\bibfield  {title} {\enquote {\bibinfo {title} {Magnon-drag
  {P}eltier effect in a {N}i-{C}u alloy},}\ }\href {\doibase
  10.1103/PhysRevB.13.2072} {\bibfield  {journal} {\bibinfo  {journal} {Phys.
  Rev. B}\ }\textbf {\bibinfo {volume} {13}},\ \bibinfo {pages} {2072--2079}
  (\bibinfo {year} {1976})}\BibitemShut {NoStop}%
\bibitem [{\citenamefont {Miura}\ and\ \citenamefont
  {Sakuma}(2012)}]{Miura:2012}%
  \BibitemOpen
  \bibfield  {author} {\bibinfo {author} {\bibfnamefont {Daisuke}\ \bibnamefont
  {Miura}}\ and\ \bibinfo {author} {\bibfnamefont {Akimasa}\ \bibnamefont
  {Sakuma}},\ }\bibfield  {title} {\enquote {\bibinfo {title} {Microscopic
  theory of magnon-drag thermoelectric transport in ferromagnetic metals},}\
  }\href {\doibase 10.1143/JPSJ.81.113602} {\bibfield  {journal} {\bibinfo
  {journal} {Journal of the Physical Society of Japan}\ }\textbf {\bibinfo
  {volume} {81}},\ \bibinfo {pages} {113602} (\bibinfo {year}
  {2012})}\BibitemShut {NoStop}%
\bibitem [{\citenamefont {Liu}\ \emph {et~al.}(2016)\citenamefont {Liu},
  \citenamefont {Vignale},\ and\ \citenamefont {Flatt\'e}}]{Flatte:2016}%
  \BibitemOpen
  \bibfield  {author} {\bibinfo {author} {\bibfnamefont {Tianyu}\ \bibnamefont
  {Liu}}, \bibinfo {author} {\bibfnamefont {G.}~\bibnamefont {Vignale}}, \ and\
  \bibinfo {author} {\bibfnamefont {Michael~E.}\ \bibnamefont {Flatt\'e}},\
  }\bibfield  {title} {\enquote {\bibinfo {title} {Nonlocal drag of magnons in
  a ferromagnetic bilayer},}\ }\href {\doibase 10.1103/PhysRevLett.116.237202}
  {\bibfield  {journal} {\bibinfo  {journal} {Phys. Rev. Lett.}\ }\textbf
  {\bibinfo {volume} {116}},\ \bibinfo {pages} {237202} (\bibinfo {year}
  {2016})}\BibitemShut {NoStop}%
\bibitem [{\citenamefont {Gangadharaiah}\ \emph {et~al.}(2010)\citenamefont
  {Gangadharaiah}, \citenamefont {Chernyshev},\ and\ \citenamefont
  {Brenig}}]{Chernyshev:2010}%
  \BibitemOpen
  \bibfield  {author} {\bibinfo {author} {\bibfnamefont {Suhas}\ \bibnamefont
  {Gangadharaiah}}, \bibinfo {author} {\bibfnamefont {A.~L.}\ \bibnamefont
  {Chernyshev}}, \ and\ \bibinfo {author} {\bibfnamefont {Wolfram}\
  \bibnamefont {Brenig}},\ }\bibfield  {title} {\enquote {\bibinfo {title}
  {Thermal drag revisited: Boltzmann versus {K}ubo},}\ }\href {\doibase
  10.1103/PhysRevB.82.134421} {\bibfield  {journal} {\bibinfo  {journal} {Phys.
  Rev. B}\ }\textbf {\bibinfo {volume} {82}},\ \bibinfo {pages} {134421}
  (\bibinfo {year} {2010})}\BibitemShut {NoStop}%
\bibitem [{\citenamefont {Lucas}\ and\ \citenamefont
  {Fong}(2018)}]{Lucas:2018}%
  \BibitemOpen
  \bibfield  {author} {\bibinfo {author} {\bibfnamefont {Andrew}\ \bibnamefont
  {Lucas}}\ and\ \bibinfo {author} {\bibfnamefont {Kin~Chung}\ \bibnamefont
  {Fong}},\ }\bibfield  {title} {\enquote {\bibinfo {title} {Hydrodynamics of
  electrons in graphene},}\ }\href {\doibase 10.1088/1361-648x/aaa274}
  {\bibfield  {journal} {\bibinfo  {journal} {Journal of Physics: Condensed
  Matter}\ }\textbf {\bibinfo {volume} {30}},\ \bibinfo {pages} {053001}
  (\bibinfo {year} {2018})}\BibitemShut {NoStop}%
\bibitem [{\citenamefont {Levchenko}\ and\ \citenamefont
  {Schmalian}(2020)}]{Levchenko:2020}%
  \BibitemOpen
  \bibfield  {author} {\bibinfo {author} {\bibfnamefont {Alex}\ \bibnamefont
  {Levchenko}}\ and\ \bibinfo {author} {\bibfnamefont {J{\"o}rg}\ \bibnamefont
  {Schmalian}},\ }\bibfield  {title} {\enquote {\bibinfo {title} {Transport
  properties of strongly coupled electron--phonon liquids},}\ }\href {\doibase
  https://doi.org/10.1016/j.aop.2020.168218} {\bibfield  {journal} {\bibinfo
  {journal} {Annals of Physics}\ }\textbf {\bibinfo {volume} {419}},\ \bibinfo
  {pages} {168218} (\bibinfo {year} {2020})}\BibitemShut {NoStop}%
\bibitem [{\citenamefont {Narozhny}(2022)}]{Narozhny:2022}%
  \BibitemOpen
  \bibfield  {author} {\bibinfo {author} {\bibfnamefont {Boris~N.}\
  \bibnamefont {Narozhny}},\ }\bibfield  {title} {\enquote {\bibinfo {title}
  {Hydrodynamic approach to two-dimensional electron systems},}\ }\href
  {\doibase 10.1007/s40766-022-00036-z} {\bibfield  {journal} {\bibinfo
  {journal} {La Rivista del Nuovo Cimento}\ }\textbf {\bibinfo {volume} {45}},\
  \bibinfo {pages} {661--736} (\bibinfo {year} {2022})}\BibitemShut {NoStop}%
\bibitem [{\citenamefont {Crossno}\ \emph {et~al.}(2016)\citenamefont
  {Crossno}, \citenamefont {Shi}, \citenamefont {Wang}, \citenamefont {Liu},
  \citenamefont {Harzheim}, \citenamefont {Lucas}, \citenamefont {Sachdev},
  \citenamefont {Kim}, \citenamefont {Taniguchi}, \citenamefont {Watanabe},
  \citenamefont {Ohki},\ and\ \citenamefont {Fong}}]{Crossno:2016}%
  \BibitemOpen
  \bibfield  {author} {\bibinfo {author} {\bibfnamefont {Jesse}\ \bibnamefont
  {Crossno}}, \bibinfo {author} {\bibfnamefont {Jing~K.}\ \bibnamefont {Shi}},
  \bibinfo {author} {\bibfnamefont {Ke}~\bibnamefont {Wang}}, \bibinfo {author}
  {\bibfnamefont {Xiaomeng}\ \bibnamefont {Liu}}, \bibinfo {author}
  {\bibfnamefont {Achim}\ \bibnamefont {Harzheim}}, \bibinfo {author}
  {\bibfnamefont {Andrew}\ \bibnamefont {Lucas}}, \bibinfo {author}
  {\bibfnamefont {Subir}\ \bibnamefont {Sachdev}}, \bibinfo {author}
  {\bibfnamefont {Philip}\ \bibnamefont {Kim}}, \bibinfo {author}
  {\bibfnamefont {Takashi}\ \bibnamefont {Taniguchi}}, \bibinfo {author}
  {\bibfnamefont {Kenji}\ \bibnamefont {Watanabe}}, \bibinfo {author}
  {\bibfnamefont {Thomas~A.}\ \bibnamefont {Ohki}}, \ and\ \bibinfo {author}
  {\bibfnamefont {Kin~Chung}\ \bibnamefont {Fong}},\ }\bibfield  {title}
  {\enquote {\bibinfo {title} {Observation of the {D}irac fluid and the
  breakdown of the {W}iedemann-{F}ranz law in graphene},}\ }\href {\doibase
  10.1126/science.aad0343} {\bibfield  {journal} {\bibinfo  {journal}
  {Science}\ }\textbf {\bibinfo {volume} {351}},\ \bibinfo {pages} {1058--1061}
  (\bibinfo {year} {2016})}\BibitemShut {NoStop}%
\bibitem [{\citenamefont {Waissman}\ \emph {et~al.}(2022)\citenamefont
  {Waissman}, \citenamefont {Anderson}, \citenamefont {Talanov}, \citenamefont
  {Yan}, \citenamefont {Shin}, \citenamefont {Najafabadi}, \citenamefont
  {Rezaee}, \citenamefont {Feng}, \citenamefont {Nocera}, \citenamefont
  {Taniguchi}, \citenamefont {Watanabe}, \citenamefont {Skinner}, \citenamefont
  {Matveev},\ and\ \citenamefont {Kim}}]{Waissman:2022}%
  \BibitemOpen
  \bibfield  {author} {\bibinfo {author} {\bibfnamefont {Jonah}\ \bibnamefont
  {Waissman}}, \bibinfo {author} {\bibfnamefont {Laurel~E.}\ \bibnamefont
  {Anderson}}, \bibinfo {author} {\bibfnamefont {Artem~V.}\ \bibnamefont
  {Talanov}}, \bibinfo {author} {\bibfnamefont {Zhongying}\ \bibnamefont
  {Yan}}, \bibinfo {author} {\bibfnamefont {Young~J.}\ \bibnamefont {Shin}},
  \bibinfo {author} {\bibfnamefont {Danial~H.}\ \bibnamefont {Najafabadi}},
  \bibinfo {author} {\bibfnamefont {Mehdi}\ \bibnamefont {Rezaee}}, \bibinfo
  {author} {\bibfnamefont {Xiaowen}\ \bibnamefont {Feng}}, \bibinfo {author}
  {\bibfnamefont {Daniel~G.}\ \bibnamefont {Nocera}}, \bibinfo {author}
  {\bibfnamefont {Takashi}\ \bibnamefont {Taniguchi}}, \bibinfo {author}
  {\bibfnamefont {Kenji}\ \bibnamefont {Watanabe}}, \bibinfo {author}
  {\bibfnamefont {Brian}\ \bibnamefont {Skinner}}, \bibinfo {author}
  {\bibfnamefont {Konstantin~A.}\ \bibnamefont {Matveev}}, \ and\ \bibinfo
  {author} {\bibfnamefont {Philip}\ \bibnamefont {Kim}},\ }\bibfield  {title}
  {\enquote {\bibinfo {title} {Electronic thermal transport measurement in
  low-dimensional materials with graphene non-local noise thermometry},}\
  }\href {\doibase 10.1038/s41565-021-01015-x} {\bibfield  {journal} {\bibinfo
  {journal} {Nature Nanotechnology}\ }\textbf {\bibinfo {volume} {17}},\
  \bibinfo {pages} {166--173} (\bibinfo {year} {2022})}\BibitemShut {NoStop}%
\bibitem [{\citenamefont {Talanov}\ \emph {et~al.}(2024)\citenamefont
  {Talanov}, \citenamefont {Waissman}, \citenamefont {Hui}, \citenamefont
  {Skinner}, \citenamefont {Watanabe}, \citenamefont {Taniguchi},\ and\
  \citenamefont {Kim}}]{Talanov:2024}%
  \BibitemOpen
  \bibfield  {author} {\bibinfo {author} {\bibfnamefont {Artem}\ \bibnamefont
  {Talanov}}, \bibinfo {author} {\bibfnamefont {Jonah}\ \bibnamefont
  {Waissman}}, \bibinfo {author} {\bibfnamefont {Aaron}\ \bibnamefont {Hui}},
  \bibinfo {author} {\bibfnamefont {Brian}\ \bibnamefont {Skinner}}, \bibinfo
  {author} {\bibfnamefont {Kenji}\ \bibnamefont {Watanabe}}, \bibinfo {author}
  {\bibfnamefont {Takashi}\ \bibnamefont {Taniguchi}}, \ and\ \bibinfo {author}
  {\bibfnamefont {Philip}\ \bibnamefont {Kim}},\ }\href
  {https://arxiv.org/abs/2406.13799} {\enquote {\bibinfo {title} {Observation
  of electronic viscous dissipation in graphene magneto-thermal transport},}\ }
  (\bibinfo {year} {2024}),\ \Eprint {http://arxiv.org/abs/2406.13799}
  {arXiv:2406.13799 [cond-mat.mes-hall]} \BibitemShut {NoStop}%
\bibitem [{\citenamefont {Gurzhi}(1968)}]{Gurzhi:1968}%
  \BibitemOpen
  \bibfield  {author} {\bibinfo {author} {\bibfnamefont {R.~N.}\ \bibnamefont
  {Gurzhi}},\ }\bibfield  {title} {\enquote {\bibinfo {title} {Hydrodynamic
  effects in solids at low temperature},}\ }\href@noop {} {\bibfield  {journal}
  {\bibinfo  {journal} {Soviet Physics Uspekhi}\ }\textbf {\bibinfo {volume}
  {11}},\ \bibinfo {pages} {255} (\bibinfo {year} {1968})}\BibitemShut
  {NoStop}%
\bibitem [{\citenamefont {Andreev}\ \emph {et~al.}(2011)\citenamefont
  {Andreev}, \citenamefont {Kivelson},\ and\ \citenamefont {Spivak}}]{AKS}%
  \BibitemOpen
  \bibfield  {author} {\bibinfo {author} {\bibfnamefont {A.~V.}\ \bibnamefont
  {Andreev}}, \bibinfo {author} {\bibfnamefont {Steven~A.}\ \bibnamefont
  {Kivelson}}, \ and\ \bibinfo {author} {\bibfnamefont {B.}~\bibnamefont
  {Spivak}},\ }\bibfield  {title} {\enquote {\bibinfo {title} {Hydrodynamic
  description of transport in strongly correlated electron systems},}\ }\href
  {\doibase 10.1103/PhysRevLett.106.256804} {\bibfield  {journal} {\bibinfo
  {journal} {Phys. Rev. Lett.}\ }\textbf {\bibinfo {volume} {106}},\ \bibinfo
  {pages} {256804} (\bibinfo {year} {2011})}\BibitemShut {NoStop}%
\bibitem [{\citenamefont {Dr{\"o}geler}\ \emph {et~al.}(2014)\citenamefont
  {Dr{\"o}geler}, \citenamefont {Volmer}, \citenamefont {Wolter}, \citenamefont
  {Terr{\'e}s}, \citenamefont {Watanabe}, \citenamefont {Taniguchi},
  \citenamefont {G{\"u}ntherodt}, \citenamefont {Stampfer},\ and\ \citenamefont
  {Beschoten}}]{Drogeler:2014}%
  \BibitemOpen
  \bibfield  {author} {\bibinfo {author} {\bibfnamefont {Marc}\ \bibnamefont
  {Dr{\"o}geler}}, \bibinfo {author} {\bibfnamefont {Frank}\ \bibnamefont
  {Volmer}}, \bibinfo {author} {\bibfnamefont {Maik}\ \bibnamefont {Wolter}},
  \bibinfo {author} {\bibfnamefont {Bernat}\ \bibnamefont {Terr{\'e}s}},
  \bibinfo {author} {\bibfnamefont {Kenji}\ \bibnamefont {Watanabe}}, \bibinfo
  {author} {\bibfnamefont {Takashi}\ \bibnamefont {Taniguchi}}, \bibinfo
  {author} {\bibfnamefont {Gernot}\ \bibnamefont {G{\"u}ntherodt}}, \bibinfo
  {author} {\bibfnamefont {Christoph}\ \bibnamefont {Stampfer}}, \ and\
  \bibinfo {author} {\bibfnamefont {Bernd}\ \bibnamefont {Beschoten}},\
  }\bibfield  {title} {\enquote {\bibinfo {title} {Nanosecond spin lifetimes in
  single- and few-layer graphene--hbn heterostructures at room temperature},}\
  }\bibfield  {booktitle} {\emph {\bibinfo {booktitle} {Nano Letters}},\ }\href
  {\doibase 10.1021/nl501278c} {\bibfield  {journal} {\bibinfo  {journal} {Nano
  Letters}\ }\textbf {\bibinfo {volume} {14}},\ \bibinfo {pages} {6050--6055}
  (\bibinfo {year} {2014})}\BibitemShut {NoStop}%
\bibitem [{\citenamefont {Dr{\"o}geler}\ \emph {et~al.}(2016)\citenamefont
  {Dr{\"o}geler}, \citenamefont {Franzen}, \citenamefont {Volmer},
  \citenamefont {Pohlmann}, \citenamefont {Banszerus}, \citenamefont {Wolter},
  \citenamefont {Watanabe}, \citenamefont {Taniguchi}, \citenamefont
  {Stampfer},\ and\ \citenamefont {Beschoten}}]{Drogeler:2016}%
  \BibitemOpen
  \bibfield  {author} {\bibinfo {author} {\bibfnamefont {Marc}\ \bibnamefont
  {Dr{\"o}geler}}, \bibinfo {author} {\bibfnamefont {Christopher}\ \bibnamefont
  {Franzen}}, \bibinfo {author} {\bibfnamefont {Frank}\ \bibnamefont {Volmer}},
  \bibinfo {author} {\bibfnamefont {Tobias}\ \bibnamefont {Pohlmann}}, \bibinfo
  {author} {\bibfnamefont {Luca}\ \bibnamefont {Banszerus}}, \bibinfo {author}
  {\bibfnamefont {Maik}\ \bibnamefont {Wolter}}, \bibinfo {author}
  {\bibfnamefont {Kenji}\ \bibnamefont {Watanabe}}, \bibinfo {author}
  {\bibfnamefont {Takashi}\ \bibnamefont {Taniguchi}}, \bibinfo {author}
  {\bibfnamefont {Christoph}\ \bibnamefont {Stampfer}}, \ and\ \bibinfo
  {author} {\bibfnamefont {Bernd}\ \bibnamefont {Beschoten}},\ }\bibfield
  {title} {\enquote {\bibinfo {title} {Spin lifetimes exceeding 12 ns in
  graphene nonlocal spin valve devices},}\ }\bibfield  {booktitle} {\emph
  {\bibinfo {booktitle} {Nano Letters}},\ }\href {\doibase
  10.1021/acs.nanolett.6b00497} {\bibfield  {journal} {\bibinfo  {journal}
  {Nano Letters}\ }\textbf {\bibinfo {volume} {16}},\ \bibinfo {pages}
  {3533--3539} (\bibinfo {year} {2016})}\BibitemShut {NoStop}%
\bibitem [{\citenamefont {D'Amico}\ and\ \citenamefont
  {Vignale}(2000)}]{Vignale:2000}%
  \BibitemOpen
  \bibfield  {author} {\bibinfo {author} {\bibfnamefont {Irene}\ \bibnamefont
  {D'Amico}}\ and\ \bibinfo {author} {\bibfnamefont {Giovanni}\ \bibnamefont
  {Vignale}},\ }\bibfield  {title} {\enquote {\bibinfo {title} {Theory of spin
  {C}oulomb drag in spin-polarized transport},}\ }\href {\doibase
  10.1103/PhysRevB.62.4853} {\bibfield  {journal} {\bibinfo  {journal} {Phys.
  Rev. B}\ }\textbf {\bibinfo {volume} {62}},\ \bibinfo {pages} {4853--4857}
  (\bibinfo {year} {2000})}\BibitemShut {NoStop}%
\bibitem [{\citenamefont {D'Amico}\ and\ \citenamefont
  {Vignale}(2003)}]{Vignale:2003}%
  \BibitemOpen
  \bibfield  {author} {\bibinfo {author} {\bibfnamefont {Irene}\ \bibnamefont
  {D'Amico}}\ and\ \bibinfo {author} {\bibfnamefont {Giovanni}\ \bibnamefont
  {Vignale}},\ }\bibfield  {title} {\enquote {\bibinfo {title} {Spin {C}oulomb
  drag in the two-dimensional electron liquid},}\ }\href {\doibase
  10.1103/PhysRevB.68.045307} {\bibfield  {journal} {\bibinfo  {journal} {Phys.
  Rev. B}\ }\textbf {\bibinfo {volume} {68}},\ \bibinfo {pages} {045307}
  (\bibinfo {year} {2003})}\BibitemShut {NoStop}%
\bibitem [{\citenamefont {Onsager}(1931{\natexlab{a}})}]{Onsager-I}%
  \BibitemOpen
  \bibfield  {author} {\bibinfo {author} {\bibfnamefont {Lars}\ \bibnamefont
  {Onsager}},\ }\bibfield  {title} {\enquote {\bibinfo {title} {Reciprocal
  relations in irreversible processes. {I}.}}\ }\href {\doibase
  10.1103/PhysRev.37.405} {\bibfield  {journal} {\bibinfo  {journal} {Phys.
  Rev.}\ }\textbf {\bibinfo {volume} {37}},\ \bibinfo {pages} {405--426}
  (\bibinfo {year} {1931}{\natexlab{a}})}\BibitemShut {NoStop}%
\bibitem [{\citenamefont {Onsager}(1931{\natexlab{b}})}]{Onsager-II}%
  \BibitemOpen
  \bibfield  {author} {\bibinfo {author} {\bibfnamefont {Lars}\ \bibnamefont
  {Onsager}},\ }\bibfield  {title} {\enquote {\bibinfo {title} {Reciprocal
  relations in irreversible processes. {II}.}}\ }\href {\doibase
  10.1103/PhysRev.38.2265} {\bibfield  {journal} {\bibinfo  {journal} {Phys.
  Rev.}\ }\textbf {\bibinfo {volume} {38}},\ \bibinfo {pages} {2265--2279}
  (\bibinfo {year} {1931}{\natexlab{b}})}\BibitemShut {NoStop}%
\bibitem [{\citenamefont {Bauer}\ \emph {et~al.}(2012)\citenamefont {Bauer},
  \citenamefont {Saitoh},\ and\ \citenamefont {van Wees}}]{Bauer:2012}%
  \BibitemOpen
  \bibfield  {author} {\bibinfo {author} {\bibfnamefont {Gerrit E.~W.}\
  \bibnamefont {Bauer}}, \bibinfo {author} {\bibfnamefont {Eiji}\ \bibnamefont
  {Saitoh}}, \ and\ \bibinfo {author} {\bibfnamefont {Bart~J.}\ \bibnamefont
  {van Wees}},\ }\bibfield  {title} {\enquote {\bibinfo {title} {Spin
  caloritronics},}\ }\href {\doibase 10.1038/nmat3301} {\bibfield  {journal}
  {\bibinfo  {journal} {Nature Materials}\ }\textbf {\bibinfo {volume} {11}},\
  \bibinfo {pages} {391--399} (\bibinfo {year} {2012})}\BibitemShut {NoStop}%
\bibitem [{\citenamefont {Landau}\ and\ \citenamefont
  {Lifshitz}(1987)}]{LL-V6}%
  \BibitemOpen
  \bibfield  {author} {\bibinfo {author} {\bibfnamefont {L.~D.}\ \bibnamefont
  {Landau}}\ and\ \bibinfo {author} {\bibfnamefont {E.~M.}\ \bibnamefont
  {Lifshitz}},\ }\href@noop {} {\emph {\bibinfo {title} {Fluid Mechanics}}},\
  \bibinfo {edition} {2nd}\ ed.,\ \bibinfo {series} {Course of Theoretical
  Physics Series}, Vol.~\bibinfo {volume} {6}\ (\bibinfo  {publisher}
  {Butterworth-Heinemann, Oxford},\ \bibinfo {year} {1987})\BibitemShut
  {NoStop}%
\bibitem [{\citenamefont {Li}\ \emph {et~al.}(2020)\citenamefont {Li},
  \citenamefont {Levchenko},\ and\ \citenamefont {Andreev}}]{Li:2020}%
  \BibitemOpen
  \bibfield  {author} {\bibinfo {author} {\bibfnamefont {Songci}\ \bibnamefont
  {Li}}, \bibinfo {author} {\bibfnamefont {Alex}\ \bibnamefont {Levchenko}}, \
  and\ \bibinfo {author} {\bibfnamefont {A.~V.}\ \bibnamefont {Andreev}},\
  }\bibfield  {title} {\enquote {\bibinfo {title} {Hydrodynamic electron
  transport near charge neutrality},}\ }\href {\doibase
  10.1103/PhysRevB.102.075305} {\bibfield  {journal} {\bibinfo  {journal}
  {Phys. Rev. B}\ }\textbf {\bibinfo {volume} {102}},\ \bibinfo {pages}
  {075305} (\bibinfo {year} {2020})}\BibitemShut {NoStop}%
\bibitem [{\citenamefont {Martin}\ \emph {et~al.}(2008)\citenamefont {Martin},
  \citenamefont {Akerman}, \citenamefont {Ulbricht}, \citenamefont {Lohmann},
  \citenamefont {Smet}, \citenamefont {von Klitzing},\ and\ \citenamefont
  {Yacoby}}]{Yacoby:2008}%
  \BibitemOpen
  \bibfield  {author} {\bibinfo {author} {\bibfnamefont {J.}~\bibnamefont
  {Martin}}, \bibinfo {author} {\bibfnamefont {N.}~\bibnamefont {Akerman}},
  \bibinfo {author} {\bibfnamefont {G.}~\bibnamefont {Ulbricht}}, \bibinfo
  {author} {\bibfnamefont {T.}~\bibnamefont {Lohmann}}, \bibinfo {author}
  {\bibfnamefont {J.~H.}\ \bibnamefont {Smet}}, \bibinfo {author}
  {\bibfnamefont {K.}~\bibnamefont {von Klitzing}}, \ and\ \bibinfo {author}
  {\bibfnamefont {A.}~\bibnamefont {Yacoby}},\ }\bibfield  {title} {\enquote
  {\bibinfo {title} {Observation of electron--hole puddles in graphene using a
  scanning single-electron transistor},}\ }\href {\doibase 10.1038/nphys781}
  {\bibfield  {journal} {\bibinfo  {journal} {Nature Physics}\ }\textbf
  {\bibinfo {volume} {4}},\ \bibinfo {pages} {144--148} (\bibinfo {year}
  {2008})}\BibitemShut {NoStop}%
\bibitem [{\citenamefont {Zhang}\ \emph {et~al.}(2009)\citenamefont {Zhang},
  \citenamefont {Brar}, \citenamefont {Girit}, \citenamefont {Zettl},\ and\
  \citenamefont {Crommie}}]{Crommie:2009}%
  \BibitemOpen
  \bibfield  {author} {\bibinfo {author} {\bibfnamefont {Yuanbo}\ \bibnamefont
  {Zhang}}, \bibinfo {author} {\bibfnamefont {Victor~W.}\ \bibnamefont {Brar}},
  \bibinfo {author} {\bibfnamefont {Caglar}\ \bibnamefont {Girit}}, \bibinfo
  {author} {\bibfnamefont {Alex}\ \bibnamefont {Zettl}}, \ and\ \bibinfo
  {author} {\bibfnamefont {Michael~F.}\ \bibnamefont {Crommie}},\ }\bibfield
  {title} {\enquote {\bibinfo {title} {Origin of spatial charge inhomogeneity
  in graphene},}\ }\href {\doibase 10.1038/nphys1365} {\bibfield  {journal}
  {\bibinfo  {journal} {Nature Physics}\ }\textbf {\bibinfo {volume} {5}},\
  \bibinfo {pages} {722--726} (\bibinfo {year} {2009})}\BibitemShut {NoStop}%
\bibitem [{\citenamefont {Xue}\ \emph {et~al.}(2011)\citenamefont {Xue},
  \citenamefont {Sanchez-Yamagishi}, \citenamefont {Bulmash}, \citenamefont
  {Jacquod}, \citenamefont {Deshpande}, \citenamefont {Watanabe}, \citenamefont
  {Taniguchi}, \citenamefont {Jarillo-Herrero},\ and\ \citenamefont
  {LeRoy}}]{LeRoy:2011}%
  \BibitemOpen
  \bibfield  {author} {\bibinfo {author} {\bibfnamefont {Jiamin}\ \bibnamefont
  {Xue}}, \bibinfo {author} {\bibfnamefont {Javier}\ \bibnamefont
  {Sanchez-Yamagishi}}, \bibinfo {author} {\bibfnamefont {Danny}\ \bibnamefont
  {Bulmash}}, \bibinfo {author} {\bibfnamefont {Philippe}\ \bibnamefont
  {Jacquod}}, \bibinfo {author} {\bibfnamefont {Aparna}\ \bibnamefont
  {Deshpande}}, \bibinfo {author} {\bibfnamefont {K.}~\bibnamefont {Watanabe}},
  \bibinfo {author} {\bibfnamefont {T.}~\bibnamefont {Taniguchi}}, \bibinfo
  {author} {\bibfnamefont {Pablo}\ \bibnamefont {Jarillo-Herrero}}, \ and\
  \bibinfo {author} {\bibfnamefont {Brian~J.}\ \bibnamefont {LeRoy}},\
  }\bibfield  {title} {\enquote {\bibinfo {title} {Scanning tunnelling
  microscopy and spectroscopy of ultra-flat graphene on hexagonal boron
  nitride},}\ }\href {\doibase 10.1038/nmat2968} {\bibfield  {journal}
  {\bibinfo  {journal} {Nature Materials}\ }\textbf {\bibinfo {volume} {10}},\
  \bibinfo {pages} {282--285} (\bibinfo {year} {2011})}\BibitemShut {NoStop}%
\bibitem [{\citenamefont {Lucas}\ \emph {et~al.}(2016)\citenamefont {Lucas},
  \citenamefont {Crossno}, \citenamefont {Fong}, \citenamefont {Kim},\ and\
  \citenamefont {Sachdev}}]{Lucas:2016}%
  \BibitemOpen
  \bibfield  {author} {\bibinfo {author} {\bibfnamefont {Andrew}\ \bibnamefont
  {Lucas}}, \bibinfo {author} {\bibfnamefont {Jesse}\ \bibnamefont {Crossno}},
  \bibinfo {author} {\bibfnamefont {Kin~Chung}\ \bibnamefont {Fong}}, \bibinfo
  {author} {\bibfnamefont {Philip}\ \bibnamefont {Kim}}, \ and\ \bibinfo
  {author} {\bibfnamefont {Subir}\ \bibnamefont {Sachdev}},\ }\bibfield
  {title} {\enquote {\bibinfo {title} {Transport in inhomogeneous quantum
  critical fluids and in the {D}irac fluid in graphene},}\ }\href {\doibase
  10.1103/PhysRevB.93.075426} {\bibfield  {journal} {\bibinfo  {journal} {Phys.
  Rev. B}\ }\textbf {\bibinfo {volume} {93}},\ \bibinfo {pages} {075426}
  (\bibinfo {year} {2016})}\BibitemShut {NoStop}%
\bibitem [{\citenamefont {Levchenko}\ \emph {et~al.}(2024)\citenamefont
  {Levchenko}, \citenamefont {Li},\ and\ \citenamefont
  {Andreev}}]{Levchenko:2024}%
  \BibitemOpen
  \bibfield  {author} {\bibinfo {author} {\bibfnamefont {Alex}\ \bibnamefont
  {Levchenko}}, \bibinfo {author} {\bibfnamefont {Songci}\ \bibnamefont {Li}},
  \ and\ \bibinfo {author} {\bibfnamefont {A.~V.}\ \bibnamefont {Andreev}},\
  }\bibfield  {title} {\enquote {\bibinfo {title} {Giant magnetoresistance in
  weakly disordered non-{G}alilean invariant conductors},}\ }\href {\doibase
  10.1103/PhysRevB.109.075401} {\bibfield  {journal} {\bibinfo  {journal}
  {Phys. Rev. B}\ }\textbf {\bibinfo {volume} {109}},\ \bibinfo {pages}
  {075401} (\bibinfo {year} {2024})}\BibitemShut {NoStop}%
\bibitem [{Note1()}]{Note1}%
  \BibitemOpen
  \bibinfo {note} {It has been shown in Ref.~\cite {Aleiner:2009} that for
  $e^2/\hbar v\ll 1$ the relaxation length $l_{\protect \text {rec}}$
  associated with electron-hole recombination is parametrically longer than the
  electron-electron scattering length $l_{\protect \text {ee}}$. In this case,
  the hydrodynamic description applies if the disorder correlation radius
  exceeds $l_{\protect \text {rec}}$. However, in graphene devices, where the
  interaction constant is of order unity, $l_{\protect \text {rec}}$ and
  $l_{\protect \text {ee}}$ are not expected to be parametrically different,
  and the condition $\xi \gg l_{\protect \text {ee}}$ is adequate for the
  applicability of the hydrodynamic description.}\BibitemShut {Stop}%
\bibitem [{Note2()}]{Note2}%
  \BibitemOpen
  \bibinfo {note} {See file of supplemental materials for further
  details}\BibitemShut {NoStop}%
\bibitem [{\citenamefont {Hui}\ \emph {et~al.}(2024)\citenamefont {Hui},
  \citenamefont {Pozderac},\ and\ \citenamefont {Skinner}}]{Skinner:2024}%
  \BibitemOpen
  \bibfield  {author} {\bibinfo {author} {\bibfnamefont {Aaron}\ \bibnamefont
  {Hui}}, \bibinfo {author} {\bibfnamefont {Calvin}\ \bibnamefont {Pozderac}},
  \ and\ \bibinfo {author} {\bibfnamefont {Brian}\ \bibnamefont {Skinner}},\
  }\bibfield  {title} {\enquote {\bibinfo {title} {Two-dimensional hydrodynamic
  electron flow through periodic and random potentials},}\ }\href {\doibase
  10.1103/PhysRevB.109.155145} {\bibfield  {journal} {\bibinfo  {journal}
  {Phys. Rev. B}\ }\textbf {\bibinfo {volume} {109}},\ \bibinfo {pages}
  {155145} (\bibinfo {year} {2024})}\BibitemShut {NoStop}%
\bibitem [{\citenamefont {Levchenko}\ \emph {et~al.}(2010)\citenamefont
  {Levchenko}, \citenamefont {Micklitz}, \citenamefont {Rech},\ and\
  \citenamefont {Matveev}}]{Levchenko:2010}%
  \BibitemOpen
  \bibfield  {author} {\bibinfo {author} {\bibfnamefont {Alex}\ \bibnamefont
  {Levchenko}}, \bibinfo {author} {\bibfnamefont {Tobias}\ \bibnamefont
  {Micklitz}}, \bibinfo {author} {\bibfnamefont {J\'er\^ome}\ \bibnamefont
  {Rech}}, \ and\ \bibinfo {author} {\bibfnamefont {K.~A.}\ \bibnamefont
  {Matveev}},\ }\bibfield  {title} {\enquote {\bibinfo {title} {Transport in
  partially equilibrated inhomogeneous quantum wires},}\ }\href {\doibase
  10.1103/PhysRevB.82.115413} {\bibfield  {journal} {\bibinfo  {journal} {Phys.
  Rev. B}\ }\textbf {\bibinfo {volume} {82}},\ \bibinfo {pages} {115413}
  (\bibinfo {year} {2010})}\BibitemShut {NoStop}%
\bibitem [{\citenamefont {DeGottardi}\ and\ \citenamefont
  {Matveev}(2015)}]{DeGottardi:2015}%
  \BibitemOpen
  \bibfield  {author} {\bibinfo {author} {\bibfnamefont {Wade}\ \bibnamefont
  {DeGottardi}}\ and\ \bibinfo {author} {\bibfnamefont {K.~A.}\ \bibnamefont
  {Matveev}},\ }\bibfield  {title} {\enquote {\bibinfo {title} {Electrical and
  thermal transport in inhomogeneous {L}uttinger liquids},}\ }\href {\doibase
  10.1103/PhysRevLett.114.236405} {\bibfield  {journal} {\bibinfo  {journal}
  {Phys. Rev. Lett.}\ }\textbf {\bibinfo {volume} {114}},\ \bibinfo {pages}
  {236405} (\bibinfo {year} {2015})}\BibitemShut {NoStop}%
\bibitem [{\citenamefont {Xin}\ \emph {et~al.}(2023)\citenamefont {Xin},
  \citenamefont {Lourembam}, \citenamefont {Kumaravadivel}, \citenamefont
  {Kazantsev}, \citenamefont {Wu}, \citenamefont {Mullan}, \citenamefont
  {Barrier}, \citenamefont {Geim}, \citenamefont {Grigorieva}, \citenamefont
  {Mishchenko}, \citenamefont {Principi}, \citenamefont {Fal'ko}, \citenamefont
  {Ponomarenko}, \citenamefont {Geim},\ and\ \citenamefont
  {Berdyugin}}]{Berdyugin:2023}%
  \BibitemOpen
  \bibfield  {author} {\bibinfo {author} {\bibfnamefont {Na}~\bibnamefont
  {Xin}}, \bibinfo {author} {\bibfnamefont {James}\ \bibnamefont {Lourembam}},
  \bibinfo {author} {\bibfnamefont {Piranavan}\ \bibnamefont {Kumaravadivel}},
  \bibinfo {author} {\bibfnamefont {A.~E.}\ \bibnamefont {Kazantsev}}, \bibinfo
  {author} {\bibfnamefont {Zefei}\ \bibnamefont {Wu}}, \bibinfo {author}
  {\bibfnamefont {Ciaran}\ \bibnamefont {Mullan}}, \bibinfo {author}
  {\bibfnamefont {Julien}\ \bibnamefont {Barrier}}, \bibinfo {author}
  {\bibfnamefont {Alexandra~A.}\ \bibnamefont {Geim}}, \bibinfo {author}
  {\bibfnamefont {I.~V.}\ \bibnamefont {Grigorieva}}, \bibinfo {author}
  {\bibfnamefont {A.}~\bibnamefont {Mishchenko}}, \bibinfo {author}
  {\bibfnamefont {A.}~\bibnamefont {Principi}}, \bibinfo {author}
  {\bibfnamefont {V.~I.}\ \bibnamefont {Fal'ko}}, \bibinfo {author}
  {\bibfnamefont {L.~A.}\ \bibnamefont {Ponomarenko}}, \bibinfo {author}
  {\bibfnamefont {A.~K.}\ \bibnamefont {Geim}}, \ and\ \bibinfo {author}
  {\bibfnamefont {Alexey~I.}\ \bibnamefont {Berdyugin}},\ }\bibfield  {title}
  {\enquote {\bibinfo {title} {Giant magnetoresistance of {D}irac plasma in
  high-mobility graphene},}\ }\href {\doibase 10.1038/s41586-023-05807-0}
  {\bibfield  {journal} {\bibinfo  {journal} {Nature}\ }\textbf {\bibinfo
  {volume} {616}},\ \bibinfo {pages} {270--274} (\bibinfo {year}
  {2023})}\BibitemShut {NoStop}%
\bibitem [{\citenamefont {Foster}\ and\ \citenamefont
  {Aleiner}(2009)}]{Aleiner:2009}%
  \BibitemOpen
  \bibfield  {author} {\bibinfo {author} {\bibfnamefont {Matthew~S.}\
  \bibnamefont {Foster}}\ and\ \bibinfo {author} {\bibfnamefont {Igor~L.}\
  \bibnamefont {Aleiner}},\ }\bibfield  {title} {\enquote {\bibinfo {title}
  {Slow imbalance relaxation and thermoelectric transport in graphene},}\
  }\href {\doibase 10.1103/PhysRevB.79.085415} {\bibfield  {journal} {\bibinfo
  {journal} {Phys. Rev. B}\ }\textbf {\bibinfo {volume} {79}},\ \bibinfo
  {pages} {085415} (\bibinfo {year} {2009})}\BibitemShut {NoStop}%
\end{thebibliography}%

\end{document}